\magnification= 1200
%            Mat

\def\Re{I \kern-.36em R}              %  Numeri reali
\def\Co{I \kern-.66em C}              %  Numeri complessi
\def\id{1 \kern-.36em 1}              %  Identita'
\def\de{\partial}                     %  Derivata parziale
\def\rdot{{\raise.5ex \hbox{.}}}      %  Moltiplicazione 
                 %  Infinito
                   %  Intersezione
                  %  Unione disgiunta
                   %  Unione

%            Frecce

\def\arr{\rightarrow }                %  Freccia da applicazione
   %  Coimplicazione con indent
      %  Implicazione con indent
\def\bullet{\vrule height4pt width4pt depth0pt} 
                                      %  Quadratino fine dimostrazione

%             Diagrammi

\def\mapright#1{\smash{\mathop{\longrightarrow}\limits^{#1}}}

\def\mapdown#1{\Big\downarrow\rlap
 {$\vcenter{\hbox{$\scriptstyle#1$}}$}}

%             Geometria

\def\fprod#1#2{#1{\mathop{\times}_M} #2}
\def\lag{{\cal L}}
\def\lagAst{\hat{\cal L}}

\def\la{\lambda}
\def\ga{\gamma}
\def\na{\nabla}

% File di definizioni GR92.STY

\font\tit=cmbx12 scaled\magstep 1
\font\abst=cmsl9

\hsize=6truein
\vsize=8.5truein
\newcount\who
\parskip=0pt

\def\centra#1{\vbox{\rightskip=0pt plus1fill\leftskip=0pt plus1fill #1}}

\def\title#1{\baselineskip=20truept\parindent=0pt\centra{\tit #1}
\medskip\baselineskip=12pt\centra{by}\def\titolo{#1}}

\def\short#1{\def\titolo{#1}}

\def\authors#1{\medskip\centra{#1}}
\long\def\addresses#1{\medskip\centra{#1}}
\long\def\addr#1#2{$^{#1}${\it #2}\par}
\long\def\support#1#2{\footnote{}{\hbox to 15truept{\hfill$^{#1}$\ }\sl #2}}
\long\def\summary#1{
\vbox{\par\leftskip=30truept\rightskip=30truept
\noindent{\bf Summary:} \abst #1}\parindent=15truept}

\def\section#1{\vskip 20pt{\bf\noindent #1}\nobreak\medskip\nobreak}

\long\def\theo#1#2{\parindent=20pt \item{}{\bf #1\abst #2}\parindent=10pt}

\long\def\dim#1{\parindent=20pt \item{}{#1}\parindent=10pt}

\newcount\firstp
\firstp=\pageno

\def\nP2{{\bf [1]}}
\def\titP2{
L. Fatibene, M. Ferraris, M. Francaviglia,
{\it N\"other Formalism for Conserved Quantities
in Classical Gauge Field Theories.
Part II: the Case of Arbitrary Bosonic Matter.}
(submitted for publication)}

\def\nHae{{\bf [2]}}
\def\titHae{
A. Haefliger, 
Comptes Rendus Acad. Sc. Paris, {\bf 243} (1956) 558--560}

\def\nBorHi{{\bf [3]}}
\def\titBorHi{
A. Borel, F. Hirzebruch,
Amer. J. Math. {\bf 80} (1958) 458--538;
Amer. J. Math. {\bf 81} (1959) 315--382;
Amer. J. Math. {\bf 82} (1960) 491--504}

\def\nMil{{\bf [4]}}
\def\titMil{
J. Milnor,
Enseignement Math. (2) {\bf 9} (1963) 198--203}

\def\nLawMi{{\bf [5]}}
\def\titLawMi{
H. Blaine Lawson Jr., M.-L. Michelsohn,
{\it Spin Geometry},
(1989) Princeton University Press, New Jersey.}

\def\nGP{{\bf [6]}}
\def\titGP{W. Greub, H.R. Petry,
{\it On the lifting of structure groups}, 
in: Lecture Notes in Mathematics
{\bf 676} (1978), 217, Springer--Verlag, NY.}

\def\nPenone{{\bf [7]}}
\def\titPenone{R. Penrose, W. Rindler,
{\it Spinors and space-time}, volume 1,
(1984), CUP, Cambridge}

\def\nOla{{\bf [8]}}
\def\titOla{B.M. van den Heuvel,
J. Math. Phys. {\bf 35} (4) (1994) 1668--1687}

\def\nGe{{\bf [9]}}
\def\titGe{R. Geroch,
J. Math. Phys. {\bf 9} (11) (1968) 1739--1744}

\def\nBW{{\bf [10]}}
\def\titBW{R.W. Bass, L. Witten,
Rev. Mod. Phys.
{\bf 29} (1957) 452--453}

\def\nPentwo{{\bf [11]}}
\def\titPentwo{R. Penrose,
{\it Structure of space-time}, 
in {\it Battelle Rencontres, 1967 Lectures in Mathematics and Physics},
eds: C.M. DeWitt and J.A. Wheeler
(1968), Benjamin, New York.}

\def\nHE{{\bf [12]}}
\def\titHE{S.W. Hawking, G.F.R. Ellis,
{\it The Large Structure of Space-Time}, 
(1973), CUP, Cambridge.}

\def\nF3G{{\bf [13]}}
\def\titF3G{L. Fatibene, M. Ferraris, M. Francaviglia, M. Godina,
{\it A geometric definition of Lie derivative for Spinor Fields},
in: Proceedings of 
{\it ``6th International Conference on Differential Geometry
and its Applications, August 28--September 1, 1995"}, (Brno, Czech Republic),
Editor: I. Kol{\'a}{\v r}, MU University, Brno, Czech Republic (1996)}

\def\nKos{{\bf [14]}}
\def\titKos{Y. Kosmann,
Ann. di Matematica Pura e Appl.
{\bf 91} (1972) 317--395}

\def\nRoot{{\bf [15]}}
\def\titRoot{
M. Ferraris, M. Francaviglia,
{\it The Lagrangian Approach to Conserved Quantities in General Relativity},
in: {\it Mechanics, Analysis and Geometry: 200 Years after Lagrange},
Editor: M. Francaviglia, Elsevier Science Publishers B.V. (1991)}

\def\nPartI{{\bf [16]}}
\def\titPartI{L. Fatibene, M. Ferraris, M. Francaviglia,
J. Math. Phys. {\bf 35} (4) (1994) 1644--1657}

\def\nCamI{{\bf [17]}}
\def\titCamI{G. Giachetta, G. Sardanashvily,
{\it Stress--Energy--Momentum Tensor in La\-gran\-gian Field Theory.
Part I: Superpotentials.} E--print: gr--qc@xxx.lanl.gov (gr--qc/9510061)}

\def\nCamII{{\bf [18]}}
\def\titCamII{G. Giachetta, G. Sardanashvily,
{\it Stress--Energy--Momentum of Affine--Metric Gravity.
Generalized Komar Superpotentials.} 
E--print: gr--qc@xxx.lanl.gov (gr--qc/9511008)}

\def\nStress{{\bf [19]}}
\def\titStress{M. Ferraris, M. Francaviglia,
J. Math. Phys. {\bf 26} (6) (1985) 1243--1252}

\def\nKol{{\bf [20]}}
\def\titKol{I. Kol{\'a}{\v r}, P.W. Michor, J. Slov{\'a}k, (1993),
				{\it Natural Operations in Differential Geometry}, 
				Springer--Verlag, NY.}

\def\nGeo{{\bf [21]}}
\def\titGeo{M. G\"ockeler and T. Sch\"ucker,
{\it Differential Geometry, Gauge Theories and Gravity},
Cambridge University Press, New York (1987)}

\def\nGau{{\bf [22]}}
\def\titGau{D. Bleecker,
{\it Gauge Theory and Variational Principles},
Addison-Wesley Publishing Company, Massachussetts (1981)}

\def\nNatI{{\bf [23]}}
\def\titNatI{D. Krupka,
{\it Some Geometric Aspects of Variational Problems in Fibered
Manifolds}, Folia Fac. Sci. Nat. UJEP Brunensis (Physica)
(1973) {\bf 14} 1--43}

\def\nNatII{{\bf [24]}}
\def\titNatII{D. Krupka,
{\it Natural Lagrangian structures},
Semester on Diff. Geom., Sept.--Dec. 1979;
Internat. J. Theoret. Phys. {\bf 17} (1978) 359--368;
Internat. J. Theoret. Phys. {\bf 15} (1976) 949--959
}

\def\nRob{{\bf [25]}}
\def\titRob{M. Ferraris, M. Francaviglia, O. Robutti (1987). In
{\it G{\'e}om{\'e}trie et Physique}
({\it Proc. Journ{\'e}es Relativistes de Marseille 1985}),
Y. Choquet--Bruhat, B. Coll, R. Kerner and A. Lichnerowicz, Eds.
(Travaux en Cours, Hermann, Paris)
}

%\def\ver{\noindent version: 001.001.006}
%\ver
%\ 
\vskip 10pt

\title {Gauge Formalism for General Relativity\goodbreak and Fermionic Matter}
\short {Gauge Formalism for General Relativity}

\authors{\ L. Fatibene, M. Ferraris, M. Francaviglia and M. Godina}

\addresses{
 \addr {}{Dipartimento di Matematica, Universit\`a 
degli Studi di Torino,\goodbreak Via Carlo Alberto 10, 10123 Torino, Italy}
          }

\vskip 8pt
\summary{A new formalism for spinors on curved spaces is developed
in the framework of variational calculus on fibre bundles.
The theory has the same structure of a gauge theory and describes the
interaction between the gravitational field and spinors.
An appropriate gauge structure is also given
to General Relativity, replacing
the metric field with spin frames.
Finally, conserved quantities and superpotentials are calculated
under a general covariant form.
}

\voffset=1.5truecm
\topskip=1truecm

\parindent=10pt

\def\Spin{\hbox{Spin}(\eta)}
\def\SO{\hbox{SO}(\eta)}
\def\GL{\hbox{GL}}
\def\SOM#1{\hbox{SO}(M,#1)}
\def\Aut#1{\hbox{Aut}(#1)}
\def\Diff#1{\hbox{Diff}(#1)}
\def\Im{\hbox{Im}}

\section{I Introduction}

In our opinion, the history of spinor field theories
may be split into two parts.
The first part has generated a framework suitable to deal with special 
relativistic theories still used to describe Fermionic particles 
in quantum field theories;
the Poincar\'e group, i.e. the isometry 
group of the Minkowski space, plays an important role in it,
so that on general curved spaces it is hard 
to build a theory in a simple and 
unconditioned way as it happens instead in the case of tensor ({\it Bosonic})
matter (see \nP2).

The second research trend has thence come up to deal with curved spaces, setting a first
step towards a general relativistic theory. Solutions to the problem should be
unrestricted or, at least, reasonably general to include all matterfields which have a
physical relevance and all admissible space-times, preferably without any stringent
symmetry requirement. A number of possible approaches have been proposed in the
literature (see \nHae, \nBorHi, \nMil, \nLawMi). Most of them rely on the following
definition:

\vskip 10pt\theo{Definition (1.1):}
{
Let $(M,g)$ be a (pseudo)-Riemannian orientable manifold; a {\it spin 
structure} on $(M,g)$ is a pair $(\Sigma,\bar\Lambda)$ where $\Sigma$ is a 
principal fibre bundle with $\Spin$ as structure group, $\eta$ being the 
signature of $g$, and $\bar\Lambda:\Sigma\arr SO(M,g)$ is a bundle morphism
such that:
\vskip 10pt
$$
  \matrix{
          \Sigma    &\mapright{\bar\Lambda}& \SOM{g}  \cr
          \mapdown{}&                        &\mapdown{}\cr
          M         &\mapright{id_M}         & M        \cr
        }
  \qquad
  \matrix{
          \Sigma                   &\mapright{R_S}& \Sigma                \cr
          \mapdown{\bar\Lambda}  &              &\mapdown{\bar\Lambda}\cr
          \SOM{g}    &\mapright{R_{\Lambda(S)}}   &\SOM{g}                \cr
        }
$$
\vskip 10pt
where $\SOM{g}$ is the $g$-orthonormal (equioriented) frame bundle,
$\Lambda:\Spin\arr\SO$ is the epimorphism 
which exhibits $\Spin$  as a two-fold covering of $\SO$ and $R_S$ and $R_{\Lambda(S)}$
are the canonical right action respectively on $\Sigma$ and $\SOM{g}$.
}
\vskip 10pt
It can be shown that also $\bar\Lambda$ is an epimorphism
and moreover a two-fold covering space map.
The obstruction to the existence of spin structures on a manifold $M$ has been 
solved by Haefliger, Milnor, Greub and Petry (see \nHae, \nMil, \nGP)
by the following theorem:

\vskip 15pt\theo{Theorem (1.2):}
{
A manifold $M$ allows spin structures of signature $\eta$ if and only if is orientable, it has a
metric $g$ with  signature $\eta$ and satisfies a topological condition which amounts
to require that the second  Stiefel-Whitney class vanishes.
}

\vskip 10pt
Under these conditions, let us choose a family
of trivializations on $\SOM{g}$ and let
$g_{\alpha\beta}$ be its $\SO$-cocycle of transition functions; we
can then build a $\Spin$-cocycle $\gamma_{\alpha\beta}$ such that:
$$
\Lambda(\gamma_{\alpha\beta}(x))=g_{\alpha\beta}(x)
\qquad \forall x\in U_{\alpha}\cap U_{\beta}
\leqno{(1.3)}$$

This $\Spin$-cocycle defines a principal bundle, which will be 
called $\Sigma(M,g)$, having with $\Spin$ as fibre and $\gamma_{\alpha\beta}$
as transition functions. It also defines a morphism 
$\bar\Lambda:\Sigma(M,g)\arr\SOM{g}$ such that 
$(\Sigma(M,g),\bar\Lambda)$ is a spin structure on $M$ according to definition (1.1).

We remark that, at least in general, there is no canonical choice of 
$\gamma_{\alpha\beta}$ since more than one inequivalent
$\Spin$-cocycles fulfilling condition $(1.3)$ can exist.

At this point, most authors choose a linear representation of $\Spin$
on a suitable real or complex vector space $V$ to build a vector bundle associated to
$\Sigma(M,g)$, the sections of which are to be identified with spinor fields.

Theories of this last kind improve a lot the situation with respect to the Minkowski 
case. Still they are unsatisfactory, because of at least two reasons:

\vskip 10pt
\dim{(1.4)
They are adequate to describe spinor fields in interaction with a fixed 
gravitational field, but if we need a true relativistic theory i.e. a theory in which
also the metric is dynamical, we must as well consider deformations of the spin structure
which are related to deformations of  the metric. However in this formalism it is hard
to talk about spin structure deformations as we are actually working with a fixed
background according to (1.1).
}

\vskip 10pt\dim{(1.5)
If we are aimed to deal with physical applications we would like to cope with the problem
of  conserved quantities. To solve this problem in the most general 
situation by means of N\"other theorem, it is necessary to define Lie derivatives 
of spinor fields. As is well known (see, e.g., \nPenone) this is quite difficult because
spinors are not {\it natural  objects} so that they cannot be dragged along arbitrary
vector fields on $M$. }
\vskip 10pt

In this paper we aim to present an alternative and new viewpoint on the description of
spinor fields on a (dynamical) curved space, by means of which we believe to have
overcome both these problems and to have developed the 
tools necessary to analyse the relations between the different solutions 
proposed earlier.

Our approach relies on a new definition of spin structures which 
from the very beginning avoids any reference to a fixed metric. In most practical 
cases our definition turns out to be equivalent to the classical one but, in general, it 
requires stricter hypotheses.

\vskip 10pt\theo{Definition (1.6): }
{
Let $M$ be an orientable manifold which admits a (pseudo)-Riemannian metric. A 
{\it free spin structure} on $M$ is a pair $(\Sigma,\tilde\Lambda)$ where $\Sigma$ is
 a principal fibre bundle with $\Spin$ as structure group and 
$\tilde\Lambda:\Sigma\arr L(M)$ is a morphism such that:
\vskip 10pt
$$
  \matrix{
          \Sigma    &\mapright{\tilde\Lambda}& L(M)  \cr
          \mapdown{p}&                        &\mapdown{\pi}\cr
          M         &\mapright{id_M}         & M        \cr
        }
  \qquad
  \matrix{
          \Sigma                   &\mapright{R_S}& \Sigma                \cr
          \mapdown{\tilde\Lambda}  &              &\mapdown{\tilde\Lambda}\cr
          L(M)    &\mapright{R_{\Lambda(S)}}   & L(M)               \cr
        }
$$
\vskip 10pt
}

We stress that under this new definition $\tilde\Lambda$ is not necessarily an
epimorphism as in the case of definition (1.1) (see \nOla).

If $M$ satisfies the conditions of Greub and Petry (see theorem (1.2)) spin structures in the 
sense of (1.1) do exist for any metric $g$ and, letting $i_g:\SOM{g}\arr L(M)$ be the
canonical  immersion, $(\Sigma(M,g),i_g\circ\bar\Lambda)$ turns out to be a free spin structure 
in the sense of (1.3). Therefore, these conditions guarantee also the
existence of free spin structures.

If $M$ is parallelizable, the conditions of Greub and Petry certainly hold, and
roughly speaking there is a bijection  between free spin structures and the ordinary spin
structures (see
theorem (2.3) below); in this case we are therefore led to think about free spin
structures just as a reformulated version of ordinary spin structures.

In general, manifolds which allow spin structures need not to be 
parallelizable; however, a remarkable result by Geroch (see \nGe)
asserts that
{\sl in dimension four and signature $(1,3)$ a noncompact manifold $M$ 
admits a spin structure if and only if it is parallelizable}.

Since compact space-times are classically forbidden by causality (see, e.g., \nBW)
and globally hyperbolic space-times have necessarily a non-compact topology
$\Re\times M_3$ (again Geroch; see \nPentwo, \nHE),
Geroch theorems seem apparently to close the
question.
In {\it reasonable} space-times, in fact, the two notions defined by (1.1)
and (1.6) do practically coincide.
Let us however remark that the topological conditions of Greub and Petry are not enough 
to build a physically meaningful general relativistic field theory since, as we said, 
they do not allow to talk about deformations of spin structures in a fairly 
general way.

Moreover and more fundamentally, Fermionic theories find their justification only in
view of quantization. As is well known, quantum techniques in gravitation often require
the use of compactifications and possibly also of signature changes, so that having at
one's own disposal a definition which allows space to dynamical metrics and holds also
in these cases seems to be rather important.
%\vskip 5pt
As for the problem $(1.5)$ of conserved quantities we envisage two alternative strategies:

\vskip 15pt\dim{
(1.7)
We can define a canonical (not natural) lift which associates a vectorfield up on $\Sigma$
to each vectorfield down on the base $M$. 
This lift allows us to define the Lie derivatives of sections of
$\Sigma$ (and its associated bundles) along vectorfields of $M$ (\nF3G, \nKos).
In our opinion, this enables us to define the
energy-momentum stress tensor (see \nP2, \nRoot, \nPartI, 
\nCamI, \nCamII, \nStress).
}

\vskip 15pt\dim
{(1.8)
An alternative way is to implement a technique similar to the method used in
gauge theory to deal with conserved quantities (\nP2, \nPartI), avoiding any 
reference to Lie derivatives with respect to vectorfields on $M$ and replacing 
them with Lie derivatives with respect to projectable vectorfields on 
$\Sigma$.
In other words, we can enlarge the symmetry group by adding the {\it vertical} 
transformations to obtain all automorphisms of $\Sigma$ which are 
canonically representable on the configuration bundle, instead of using 
$\Diff{M}$ which has no such natural representation.
}
\vskip 10pt

Although both approaches seem to be viable (at least a priori),
we will herein develop the second approach.
It is in our intention, however, to investigate also the remaining approach and to
discuss and analyse its relations with other
techniques in forthcoming papers.
We remark that the second strategy we have chosen is, a
priori, the most difficult to  interpret.
In fact by enlarging the symmetry group we
add degrees of  freedom to the conserved currents and consequently we expect to have more
conserved  quantities than those we are able to interpret in our case of Fermionic
matter. However, we will see that the vertical contributions to the currents
vanish identically off-shell (owing to covariance) so that no additional conserved 
quantities will at last be defined but the energy-momentum tensors.

We would  finally like to stress that the reformulation
of General Relativity in
terms of free spin structures is in our opinion essential if we want to treat
spinor theories as  gauge theories, since in our formulation spinors interact directly
with  spin structures while ordinary Bosonic matter just interacts with the metric
associated to it.

\section{II Deformation of free spin structures and notation}

Let us choose an orientable manifold $M$ such that 
there exists on $M$ a metric $g$ of signature $\eta$.
We will call $\eta_{ab}=g(u_a,u_b)$ the metric components of $g$ with respect
to an orthonormal (local) basis.
\vskip 10pt
\theo{Assumption (2.1): }{Let $\Sigma$ be a principal bundle with fibre $\Spin$ called 
{\sl structure bundle} and let us assume that there exists at least one morphism 
$\tilde\Lambda:\Sigma\arr L(M)$ such that $(\Sigma,\tilde\Lambda)$
is a free spin structure.}
\vskip 10pt
We remark that we are not fixing $\tilde\Lambda$ uniquely; we are rather asking
that such a
 morphism exists. This condition is not always guaranteed depending on the
 topology of $M$.
For example, if $M$ is not parallelizable one cannot 
choose $\Sigma$ to be a trivial bundle on $M$. In this case, in fact, if one 
 morphism $\tilde\Lambda$ exists, then $L(M)$ should admit a global section and this is
a contradiction.
However, if $M$ satisfies the conditions of theorem $(1.2)$, one can choose a metric $g$
 and build $\Sigma(M,g)$ as explained above. This bundle allows, by
 construction, a free spin structure. As a consequence, conditions $(1.2)$ are sufficient
 conditions for the existence of at least one structure bundle $\Sigma$.

We stress that, in general, there can be more than one structure bundle 
on $M$ (see \nOla). This situation is not different from what one encounters in gauge
theories  when we fix the gauge bundle $P$ and, here as in that case, there is
no point in looking for a canonical choice; different choices give raise to 
different theories. In other words fixing $\Sigma$ is part of the
{\it system specification}.

\vskip 10pt\theo{Definition (2.2) }{
A {\it spin frame on $\Sigma$} is a morphism $\tilde\Lambda:\Sigma\arr L(M)$ for which
$(\Sigma,\tilde\Lambda)$ is a free spin structure.
}
\vskip 10pt

We are now able to state and prove the following:

\vskip 10pt\theo{Theorem (2.3)}{
Let $M$ be parallelizable.
Then there exists a bijection between spin frames and spin structures
on $\Sigma$.
Moreover, for each metric $g$ on $M$ there exists a spin frame $\tilde\Lambda$ such that
all frames in $\Im(\tilde\Lambda)\subset L(M)$ are $g$-orthonormal frames.
}
\vskip 5pt
\dim{In fact, if $M$ is parallelizable for each pair of metrics ($g$, $\tilde g$) on $M$
there exists an isomorphism $\Phi:\SOM{g}\arr\SOM{\tilde g}$
such that the following diagram commutes:

$$
  \matrix{
          \SOM{g}   & \mapright{\Phi}        & \SOM{\tilde g}  \cr
          \mapdown{}&                        &\mapdown{}\cr
          M         &\mapright{id_M}         & M        \cr
        }
\leqno{(2.4)}$$
\vskip 10pt

Accordingly, we can choose $\Sigma=\Sigma(M,g)=\Sigma(M,\tilde g)$.
Thence for each spin structure $(\Sigma,\bar\Lambda)$ we can build a spin 
frame on $\Sigma$ by composition with the canonical injection:

$$
  \matrix{
          \Sigma    & \mapright{\bar\Lambda} & \SOM{g}  \cr
\mapdown{\tilde\Lambda}&                        &\mapdown{i_g}\cr
          L(M)         &\mapright{id_M}         & L(M)        \cr
        }
\leqno{(2.5)}$$
\vskip 5pt
On the other hand, for each spinor frame
$\tilde\Lambda:\Sigma\arr L(M)$
there exists one and only one metric $\tilde g$ having the frames of 
$\Im(\tilde\Lambda)$ as orthonormal frames.
If we build $\SOM{\tilde g}$ from diagram $(2.5)$ we
infer that $\tilde\Lambda$ induces $\bar\Lambda$.
These two maps are inverse of each other, and define the bijection
as we claimed. \ \bullet
}
\vskip 10pt

To summarize, if $M$ is parallelizable we can consider spin frames instead of
spin structures on a structure bundle $\Sigma$ fixed once for all; moreover, every metric
 can be associated to some spin frame.
\vskip 5pt
If we want to consider a field theory in which spin frames are dynamical
we must first construct a fibre bundle the sections of which represent spin
frames. 

Let us then consider the following action on the manifold $\GL(m)$:
$$
\rho:(\GL(m)\times\Spin)\times \GL(m)\arr \GL(m):((A^\mu_\nu,S), e^\nu_a)
\mapsto A^\mu_\nu e^\nu_b\Lambda^b_a(S^{-1})
\leqno{(2.6)}
$$
together with the associated  bundle
$\Sigma_\rho=({\fprod{(L(M)}{\Sigma)}}\times \GL(m))/\rho$, where $m$
is the dimension of $M$.
According to the theory of gauge-natural bundles and gauge-natural operators
(see \nKol) $\fprod{L(M)}{\Sigma}$ is nothing but the principal prolongation
of the principal fibre bundle $\Sigma$, also denoted by $W^{1,0}(\Sigma)$,
with structure group $\GL(m)\times\Spin$.
It turns out that $\Sigma_\rho$ is a fibre bundle associated to
$W^{1,0}(\Sigma)$, i.e. a gauge-natural bundle of order $(1,0)$.
The bundle $\Sigma_\rho$ will be called the bundle of {\sl spin tetrads} or simply
(by the following result) the bundle of {\sl spin frames}.

\vskip 10pt\theo {Theorem (2.7) }{
Sections of $\Sigma_\rho$ are in one-to-one correspondence with spin 
frames on $\Sigma$.
}
\vskip 5pt
\dim{The bijection is the following. Let $\tilde\Lambda:\Sigma\arr L(M)$ be a
 spin frame, $\sigma^{(\alpha)}(x)$ be the identity (local) sections with 
respect to a trivialization of $\Sigma$ and 
$u^{(\alpha)}_a(x)=\tilde\Lambda(\sigma^{(\alpha)}(x))$ the corresponding (local)
sections of $L(M)$. To these objects we can associate (local) sections:

$$
s^\alpha(x)=[\sigma^{(\alpha)}(x),u^{(\alpha)}_a(x),\id]\in\Sigma_\rho
$$

\noindent which glue together to generate a global section $s$ on $\Sigma_\rho$, which is 
said to {\sl represent} our spin frame.\ \bullet}
\vskip 10pt
If we choose an automorphism $\Phi\in\Aut{\Sigma}$ of the structure bundle,
it can be represented on the bundle $\Sigma_\rho$ in the following way:

$$
\Phi_\rho:\Sigma_\rho\arr\Sigma_\rho:[u_a,p,e_a^\mu]\mapsto[L(f)(u_a),\Phi(p),e_a^\mu]
\leqno{(2.8)}
$$

\noindent where $f:M\arr M$
is the projection of $\Phi$ on $M$ and $L(f)$ is the
natural lift of $f$ to $L(M)$.
It can be easily checked that this is a good definition.

Letting $\Xi$ be the infinitesimal generator of a 1-parameter subgroup $\{\Phi_t\}$ of
automorphisms on $\Sigma$, let us denote by $\Xi_\rho$ the generator of the
subgroup induced on $\Sigma_\rho$ by $(2.8)$.
The flow of $\Xi_\rho$ drags any section of $\Sigma_\rho$, thus defining a family
$\tilde\Lambda_t$ of spin frames which will be for us, by definition, {\sl an
infinitesimal deformation of the spin frame $\tilde\Lambda_0$} (see \nOla).

Finally let us remark that, whenever we choose a trivialization of $\Sigma$ and
on $L(M)$ induced respectively by local sections $\sigma^{(\alpha)}$
and
$\de^{(\alpha)}_\mu$, we can locally choose standard representatives on $\Sigma_\rho$
as follows: $$
[\sigma^{(\alpha)},\de^{(\alpha)}_\mu,e^\mu_a]
\leqno{(2.9)}$$
so that $(x^\mu,e^\mu_a)$ are (local) coordinates in $\Sigma_\rho$.
The generic automorphism $\Phi(x,S)=(f(x),\phi(x)\rdot S)$ induces then
the following automorphism on $\Sigma_\rho$:
$$
\Phi_\rho(x^\mu,e^\mu_a)=(f^\mu(x),J^\mu_\nu e^\nu_b
\Lambda^b_a(\phi^{-1}(x)))\qquad\qquad J^\mu_\nu:=\de_\nu f^\mu(x)
\leqno{(2.10)}$$
For other standard notation see, e.g., \nRoot, \nPartI, \nGeo\ and \nGau.
%  The following 13 lines establish the use of the Euler Fraktur font.
%
\font\teneuf=eufm10 at 10pt
\font\seveneuf=eufm7 at 7pt
\font\fiveeuf=eufm5 at 5pt
\newfam\euffam
\textfont\euffam=\teneuf
\scriptfont\euffam=\seveneuf
\scriptscriptfont\euffam=\fiveeuf
\def\frak{\relaxnext@\ifmmode\let\next\frak@\else
\def\next{\Err@{Use \string\frak\space only in math mode}}\fi\next}
\def\goth{\relaxnext@\ifmmode\let\next\frak\else
\def\next{\Err@{Use \string\goth\space only in math mode}}\fi\next}
\def\frak#1{{\fam\euffam#1}}
\def\frakk#1{\noaccents@\fam\euffam#1}
%  End definition of Euler Fraktur font.

%             Formato

\def\nfor#1{\leqno(#1)}
\def\uno{1\kern-2.5pt{\rm l}}									%  Unitˆ matriciale

\section{III Covariant Lagrangians}

Let $M$ be a parallelizable and
orientable manifold which admits
(pseudo)-Rie\-man\-nian metrics of signature $\eta$. Let $\Sigma$ be
our structure bundle and
$\lambda$ be a linear representation of $\Spin$ on a suitable vector
space $V$.
We can then construct the associated vector bundle
$\Sigma_{\lambda}=:\Sigma\times_{\lambda} V$. Any
$\Phi\in\Aut\Sigma$ can be represented on $\Sigma_\la$ as follows:

$$
\Phi_{\lambda}: \Sigma_{\lambda} \arr \Sigma_{\lambda}
:[p,v]\mapsto [\Phi(p),v] \quad .
\nfor{3.1}$$

\noindent Moreover, since we aim to describe a spinor field
(not subject to any further gauge symmetry) in interaction with the
gravitational field, our {\it configuration space} will be assumed to be
the following bundle:
$$
B= \fprod{\Sigma_{\rho}}\Sigma_{\lambda}
\nfor{3.2}$$ 
and the Lagrangian will be chosen in the following form:
$$
L: \fprod{J^{2}\Sigma_{\rho}}J^{1}\Sigma_{\lambda}
\arr A^0_m(M)
\nfor{3.3}$$
According to the principle of minimal coupling, the Lagrangian $L$
is assumed to split into two parts $L=L_{H}+L_{D}$, with:
$$
L_{H}: J^{2}\Sigma_{\rho}\arr A^0_m(M)
\quad (gravitational \> Lagrangian)
\nfor{3.4}$$
$$
L_{D}: J^{1}(\fprod{\Sigma_{\rho}}\Sigma_{\lambda})\arr A^0_m(M)
\quad (spinor \> Lagrangian)
\nfor{3.5}$$

For $L_{H}$ we can take the standard Hilbert Lagrangian 
$L_H= -(1/2\kappa)\,R(j^{2}g)\sqrt{g}\,ds$ written
in metric coordinates
$g={\bar e}^{a}_{\mu}\eta_{ab}{\bar e}^{b}_{\nu}
dx^{\mu}\otimes dx^{\nu}$ where
$\| {\bar e}^{a}_{\mu}\| = \| e_{a}^{\mu} \| ^{-1}$
being $ds=dx^1\land\ldots\land dx^m$
the local volume element and $g=\vert \det \| g_{\mu\nu} \| \, \vert$.

We require that $L$ be covariant with respect to any
{\sl generalised spinor transformation}, i.e. with respect to any
element of $\Aut\Sigma$.
Since locally we have the following transformations rules:
$$
\leqalignno{
&e'^{\mu}_{a}= J^{\mu}_{\nu} e^{\nu}_{b} \Lambda^{b}_{a}(\phi^{-1}(x))
&(3.6)\cr
&
v'^A= \lambda^{A}_{B}(\phi(x))v^B
&(3.7)\cr
&
g'_{\mu\nu}= \Lambda^{a}_{c}(\phi(x)){\bar e}^{c}_{\sigma}
{\bar J}^{\sigma}_{\mu}\eta_{ab}\Lambda^{b}_{d}(\phi(x)){\bar e}^{d}_{\rho}
{\bar J}^{\rho}_{\nu}={\bar J}^{\sigma}_{\mu}g_{\sigma\rho}{\bar J}^{\rho}_{\nu}
&(3.8)\cr
}
$$
the Hilbert Lagrangian is covariant with respect to all these tranformations.
As for the rest, we have to seek conditions which must to be
satisfied by $L_{D}$.
To this purpose let us choose local coordinates in $J^{1}B$
as follows:
$$
(x^{\mu}, e^{\rho}_{a}, e^{\rho}_{a\sigma}, v^{A},
 \Omega^{A}_{a}, v^{\dagger}_{A}, \Omega^{\dagger}_{Aa})
\nfor{3.9}$$
where
$$
\Omega^{A}_{a}\equiv e^{\mu}_{a} \Omega^{A}_{\mu}
:= e^{\mu}_{a} (v^{A}_{\mu} + 
\lambda^{A}_{Bij}\Gamma^{ij}_{\mu}v^{B})
\nfor{3.10}$$
$$
\leqalignno{
&\lambda^{A}_{Bij} := {1\over 8} \partial^{\beta}_{\alpha}
\lambda^{A}_{B}(e)[\gamma_{i}, \gamma_{j}]_{\beta}^{\alpha},\qquad
\de_\alpha^\beta:={\de\over\de S^\alpha_\beta}
&(3.11)\cr
&
\Gamma^{ij}_{\mu} := {\bar e}^{i}_{\rho}
 (\Gamma^{\rho}_{\sigma\mu}e^{\sigma}_{k} + e^{\rho}_{k\mu})\eta^{kj}
&(3.12)\cr
&
\Gamma^{\rho}_{\sigma\mu} := {1\over 2} g^{\rho\nu}
(-d_{\nu}g_{\sigma\mu} + d_{\sigma}g_{\mu\nu}
 + d_{\mu}g_{\nu\sigma})
&(3.13)\cr}
$$
being $\gamma_{i}$ a set of Dirac matrices fixed
to define the two-fold covering $\Lambda : \Spin \arr \SO$.

\noindent Let us notice that, if we denote by
${\vec \sigma_{ij}}:= {1\over 8}([\gamma_{i}, \gamma_{j}]S)_{\beta}^{\alpha}
\partial^{\beta}_{\alpha}$ a system of
right-invariant vector fields over $\Spin$, the quantities
$\Gamma^{ij}_{\mu}$ defined by $(3.12)$ are nothing but the coefficients of
the spinor connection induced canonically on $\Sigma$
by the Levi--Civita connection:
$$
\omega = dx^{\mu} \otimes 
({\vec \partial}_{\mu} - \Gamma^{ij}_{\mu} {\vec \sigma_{ij}})
\nfor{3.14}$$
while the expression $\Omega^{A}_{a}$ defined by $(3.10)$
is the formal covariant derivative of the field $v^{A}$ with respect to
the connection on $\Sigma_{\lambda}$ induced by
this (principal) spinor connection.

\noindent Taking into account the transformation rule:
$$
\Gamma'^{ij}_{\mu} = {\bar J}^{\nu}_{\mu}\Lambda^{i}_{k}(\phi(x))
\Big(\Gamma^{kh}_{\nu}\Lambda^{j}_{h}(\phi(x)) + 
d_{\nu}\Lambda^{k}_{l}(\phi^{-1}(x))
\eta^{lj}\Big)
\nfor{3.15}$$
together with the identity:
$$
{(\phi^{-1}(x))^{\gamma}_{\beta}}\partial_{\mu}\phi^{\alpha}_{\gamma} =
- {1\over 8} J^{\nu}_{\mu}\Gamma'^{ij}_{\nu}
[\gamma_{i}, \gamma_{j}]^{\alpha}_{\beta}
+ {1\over 8} \Lambda^{i}_{k}(\phi(x))\Lambda^{j}_{h}(\phi(x))
\Gamma^{kh}_{\mu}[\gamma_{i}, \gamma_{j}]^{\alpha}_{\beta}
\nfor{3.16}$$
we obtain the transformation rule:
$$
\Omega'^{A}_{a} = \lambda^{A}_{B}(\phi(x)) \Omega^{B}_{b}
 \Lambda^{b}_{a}(\phi^{-1}(x))\quad .
\nfor{3.17}$$
In the chosen coordinates the Lagrangian $(3.5)$ has, in general,
the following form:
$$
L_{D}=\lag_{D}(x^{\mu}, e^{\mu}_{a}, e^{\mu}_{a\rho},
 v^{A}, v^{\dagger}_{A}, \Omega^{A}_{a}, \Omega^{\dagger}_{Aa}
)\sqrt{g}\,ds
\nfor{3.18}$$

\noindent If we require the spinor Lagrangian $L_{D}$
to be covariant with respect
to any automorphisms of the structure bundle $\Sigma$
(see \nNatI\ and \nNatII\ for the natural case)
its associated scalar density $\lagAst=\lag_{D}\sqrt{g}$ must
satisfy the following identity:
$$\leqalignno{
d_{\sigma}(\lagAst\xi^{\sigma})=&
{\partial\lagAst\over \partial v^{A}}\hbox{{\it \$}}_\Xi v^{A} + 
{\partial\lagAst\over \partial \Omega^{A}_{a}}
\hbox{{\it \$}}_\Xi \Omega^{A}_{a} + 
{\partial\lagAst\over \partial v^{\dagger}_{A}}
\hbox{{\it \$}}_\Xi v^{\dagger}_{A} + &(3.19)\cr
&+ {\partial\lagAst\over \partial \Omega^{\dagger}_{Aa}}
\hbox{{\it \$}}_\Xi \Omega^{\dagger}_{Aa} + 
{\partial\lagAst\over \partial e^{\mu}_{a}}
\hbox{{\it \$}}_\Xi e^{\mu}_{a} +
{\partial\lagAst\over \partial e^{\mu}_{a\sigma}}
\hbox{{\it \$}}_\Xi e^{\mu}_{a\sigma}\cr}$$
%\nfor{3.19}$$
where $\Xi$ is the infinitesimal generator of a one--parameter subgroup
of automorphisms of $\Sigma$ and $\xi$ is its projection on $M$.
The identity $(3.19)$ holds if and only if $\lag_{D}$ does not depend on
$(x^{\mu}, e^{\mu}_{a}, e^{\mu}_{a\sigma})$ and moreover
the following identity holds:
$$
\lambda^{A}_{Blm}\Big({\partial\lag_{D}\over \partial v^{A}}v^{B} +
{\partial\lag_{D}\over \partial \Omega^{A}_{a}}\Omega^{B}_{a}\Big) - 
{\partial\lag_{D}\over \partial \Omega^{A}_{a}}\eta_{a[m}\Omega^{A}_{l]}
+  c.c.  = 0
\nfor{3.20}$$
where $c.c.$ stands for the complex conjugate terms.

\section{IV The Dirac Lagrangian on curved spaces}

We intend here to define the generalization to curved spaces of the Dirac Lagrangian
used in Quantum Field Theory to descibe Fermionic fields.
We content ourselves to discuss the case
of Fermionic fields on a four-dimensional curved space--time $M$ which admits a metric
of Lorentzian signature $\eta=(+,-,-,-)$.
Anyhow we remark that we are not fixing a particular metric as background, but it is to
be understood as determined by the spin-tetrad field.
This generalization will also provide us an example of covariant spinor Lagrangian.

We first recall that the Dirac matrices are defined as follows:
\vskip 3pt
$$
\gamma_{0} = \Big(\matrix{0 & \uno \cr
																						\uno & 0 \cr}\Big)\quad
\gamma_{1} = \Big(\matrix{0 & \sigma_{1} \cr
																						-\sigma_{1} & 0 \cr}\Big)\quad
\gamma_{2} = \Big(\matrix{0 & \sigma_{2} \cr
																						-\sigma_{2} & 0 \cr}\Big)\quad
\gamma_{3} = \Big(\matrix{0 & \sigma_{3} \cr
																						-\sigma_{3} & 0 \cr}\Big)
\nfor{4.1}$$
\vskip 3pt\noindent
where $\sigma_{1}, \> \sigma_{2}, \> \sigma_{3}$
are the Pauli matrices defined
by: $$\sigma_{1} = \Big(\matrix{0 & 1 \cr
																								1 & 0 \cr}\Big)\qquad
  \sigma_{2} = \Big(\matrix{0 &-i \cr
																								i & 0 \cr}\Big)\qquad
		\sigma_{3} = \Big(\matrix{1 & 0 \cr
																								0 &-1 \cr}\Big)$$
The group $\Spin$ is defined to be the set of matrices
$S \in\hbox{GL}(4,\Co)$ such that an element $\Lambda \in\SO$ exists 
for which the following holds:
\vskip 3pt
$$
S\rdot \gamma_{i}\rdot S^{-1} = \Lambda^{j}_{i} \gamma_{j}
\nfor{4.2}$$
\vskip 3pt\noindent
This matrix $\Lambda \in \SO$ is by definition
the image of $S$ with respect to the homomorphism
$\Lambda : \Spin \arr \SO$.
The group $\Spin$ acts canonically on $V=\Co^{4}$ by a representation which we denote
 by $\lambda$. We can then construct
the associated vector
bundle $\Sigma_{\lambda} = \Sigma\times_{\lambda} \Co^{4}$
as in the general case explained above.
The Dirac Lagrangian is then defined as follows:
\vskip 3pt
$$
L_{D}=\Big[{i\over2}\,{\bar v}\rdot \gamma^{a}\rdot \Omega_{a} -
{i\over2}\,{\bar\Omega}_{a}\rdot\gamma^{a}\rdot v - m\,{\bar v}\rdot v\Big]
\sqrt{g}\,ds
\nfor{4.3}$$
\vskip 3pt\noindent
where ${\bar v}=v^{\dagger}\gamma_{0}\quad$,
${\bar\Omega}_{a}=\Omega^{\dagger}_{a}\gamma_{0}\quad$,
$\gamma^{a}=\eta^{ab}\gamma_{b}$ and $\,\rdot\,$ denotes the matrix product.
It is easy to verify that the Lagrangian defined by $(4.3)$
is covariant, i.e. it fulfills condition $(3.20)$.

This Lagrangian is of particular importance because,
on $M=\Re^{4}$ it reduces to the {\it standard} Dirac Lagrangian which provides us
the only spinor theory which is truly understood and experimentally tested,
and has therefore to be reproduced by any generalised spinor theory.

\section{V Conserved Quantities and Superpotentials}

We shall herein rely on N\"other theorem to generate conserved currents
associated to a family of
generalised spinor transformations  $\Phi_{t}\in\Aut\Sigma$. Let
this family be generated
by the vectorfield 
$$
\Xi = \xi^{\mu}{\vec \partial}_{\mu} + \xi^{ij}{\vec \sigma}_{ij}
$$
where
$\xi^{\mu}= {\partial \over \partial t}(f_{t})^{\mu}|_{t = 0}$
and
$\xi^{ij}= \partial_{\alpha}^{\beta}\Lambda^{i}_{k}(e)\eta^{kj}
{\partial \over \partial t}(\phi_{t})^{\alpha}_{\beta}|_{t = 0}$.
In metric coordinates $(x^{\lambda}, g_{\mu\nu})$ the Hilbert
Lagrangian reads as:
$$
L_{H}= \lag_{H}\sqrt{g}\,ds
= -{1\over{2\kappa}}\, g^{\mu\nu}R_{\mu\nu}(j^{2}g)\sqrt{g}\,ds
$$
Since both Lagrangians $L_{H}$ and $L_{D}$ are separately covariant, the following
two identities hold:
$$d_{\sigma}(\lag_{H}^{\ast}\xi^{\sigma})=
p^{\mu\nu}\hbox{{\it \$}}_\Xi g_{\mu\nu} + 
p^{\alpha\beta\gamma\delta}\hbox{{\it \$}}_\Xi R_{\alpha\beta\gamma\delta}
\nfor{5.1}$$
$$
d_{\sigma}(\lagAst\xi^{\sigma})=
p_{A}\hbox{{\it \$}}_\Xi v^{A} + 
p^{a}_{A}\hbox{{\it \$}}_\Xi \Omega^{A}_{a} + 
p^{A}\hbox{{\it \$}}_\Xi v^{\dagger}_{A} +
p^{Aa}\hbox{{\it \$}}_\Xi \Omega^{\dagger}_{Aa} + 
p^{a}_{\mu}\hbox{{\it \$}}_\Xi e^{\mu}_{a}
\nfor{5.2}$$
where we have defined the {\sl naive momenta} of $L$ by:
$$\leqalignno{
&p^{\mu\nu}={\partial\lag_{H}^{\ast}\over \partial g_{\mu\nu}},\quad
p^{\alpha\beta\gamma\delta}={\partial\lag_{H}^{\ast}\over \partial
R_{\alpha\beta\gamma\delta}},\quad
p_{A}={\partial\lagAst\over \partial v^{A}},\quad
p^{A}={\partial\lagAst\over \partial v^{\dagger}_{A}},&(5.3)\cr
&p^{a}_{A}={\partial\lagAst\over \partial \Omega^{A}_{a}},\quad
p^{Aa}={\partial\lagAst\over \partial \Omega^{\dagger}_{Aa}},\quad
p^{a}_{\mu}={\partial\lagAst\over \partial e^{\mu}_{a}},\quad
\tilde p^{\alpha\beta\gamma\delta}=p^{\alpha ( \beta\gamma ) \delta}. \cr}$$
%\nfor{5.3}$$
Starting from $(5.1)$, substituting the expressions
of the Appendix and integrating covariantly by parts, we are finally led to
the following formulas for the {\sl gravitational current} and {\sl work}:
$$
\nabla_{\lambda}E^{\lambda}(L_{H},\Xi)=W(L_{H},\Xi)
\nfor{5.4}$$
$$\leqalignno{
E^{\lambda}(L_{H},\Xi) &=
{\buildrel H \over {T}}\hbox{$^{\lambda}_{\sigma}\xi^{\sigma}$} 
+ {\buildrel H \over {T}}\hbox{$^{\lambda\mu}_{\sigma}\nabla_{\mu}\xi^{\sigma}$} 
+ {\buildrel H \over {T}}
\hbox{$^{\lambda\mu\nu}_{\sigma}\nabla_{\mu\nu}\xi^{\sigma}$}=&(5.5)\cr
&= -4 \tilde p^{\theta\mu\nu\lambda}{\bar e}_{\mu}^{a} g_{\nu\sigma}
\nabla_{\theta}\hbox{{\it \$}}_\Xi e^{\sigma}_{a} +
4 \nabla_{\theta}\tilde p^{\lambda\mu\nu\theta}{\bar e}_{\mu}^{a} g_{\nu\sigma}
\hbox{{\it \$}}_\Xi e^{\sigma}_{a} - 
\lag_{H}^{\ast}\xi^{\lambda}\cr
W(L_{H},\Xi) &= 2 e^{\mu\nu}(L_{H}){\bar e}_{\mu}^{a} g_{\nu\sigma}
\hbox{{\it \$}}_\Xi e^{\sigma}_{a} = 
{\buildrel H \over {W}}\hbox{$^{\mu}_{\sigma}\nabla_{\mu}\xi^{\sigma}$}
&(5.6)\cr
}$$
with 
$\nabla_{\mu\nu}\xi^{\sigma}={1\over2}
(\nabla_\mu\nabla_\nu+\nabla_\nu\nabla_\mu)\xi^{\sigma}$
and:
$$
\leqalignno{
e^{\mu\nu}(L_{H}) &= p^{\mu\nu} + p^{(\mu\vert\beta\gamma\delta\vert}
R^{\nu)}_{\beta\gamma\delta} + 
2\nabla_{\lambda}\nabla_{\theta}\tilde p^{\lambda\mu\nu\theta}
&(5.7)\cr
{\buildrel H \over {W}}\hbox{$^{\mu}_{\sigma}$}&=
- 2 e^{\mu\nu}(L_{H}) g_{\nu\sigma}
&(5.8)\cr
{\buildrel H \over {T}}\hbox{$^{\lambda}_{\sigma}$} &= 
- 2 \tilde p^{\nu\mu\theta\lambda}R_{\sigma\theta\nu\mu} - 
\lag_{H}^{\ast}\delta^{\lambda}_{\sigma}
&(5.9)\cr
{\buildrel H \over {T}}\hbox{$^{\lambda\mu}_{\sigma}$} &= 
-4 \nabla_{\theta}\tilde p^{\lambda\mu\nu\theta} g_{\nu\sigma}\equiv 0
&(5.10)\cr
{\buildrel H \over {T}}\hbox{$^{\lambda\mu\nu}_{\sigma}$} &= 
- 2 \tilde p^{\rho\mu\nu\lambda}g_{\rho\sigma}
&(5.11)\cr}
$$
As a consequence, the covariant conditions for the gravitational part are:

$$
\cases{
\nabla_{\lambda}{\buildrel H \over {T}}\hbox{$^{\lambda}_{\rho}$} + 
{1 \over 2}\,{\buildrel H \over {T}}\hbox{$^{\nu\mu}_{\sigma}$}
R^{\sigma}_{\rho\nu\mu} +
{1 \over 3}\,\nabla_{\mu} R^{\sigma}_{\rho\lambda\nu} 
{\buildrel H \over {T}}\hbox{$^{\lambda\nu\mu}_{\sigma}$} = 0 \cr
{\buildrel H \over {T}}\hbox{$^{\mu}_{\rho}$} + 
\nabla_{\lambda}{\buildrel H \over {T}}\hbox{$^{\lambda\mu}_{\rho}$} +
{\buildrel H \over {T}}\hbox{$^{\lambda\nu\mu}_{\sigma}$}
R^{\sigma}_{\rho\lambda\nu} -
{2 \over 3}\,
{\buildrel H \over {T}}\hbox{$^{\lambda\sigma\nu}_{\rho}$}
R^{\mu}_{\sigma\lambda\nu} = 
{\buildrel H \over {W}}\hbox{$^{\mu}_{\rho}$} \cr
{\buildrel H \over {T}}\hbox{$^{(\mu\nu)}_{\rho}$} +
\nabla_{\lambda}{\buildrel H \over {T}}\hbox{$^{\lambda\mu\nu}_{\rho}$} = 0 \cr
{\buildrel H \over {T}}\hbox{$^{(\lambda\mu\nu)}_{\rho}$} = 0 \cr
}\nfor{5.12}$$

\noindent We remark that the contribution to the vertical current vanishes
identically.

Taking now $(5.2)$ into account, substituting the expressions
of the Appendix and integrating again (covariantly) by parts, we obtain the following:
$$
\leqalignno{
\nabla_{\lambda}E^{\lambda}(L_{D},\Xi)&=W(L_{D},\Xi)
&(5.13)\cr
E^{\lambda}(L_{D},\Xi) &= 
{\buildrel D \over {T}}\hbox{$^{\lambda}_{\sigma}\xi^{\sigma}$} 
+ {\buildrel D \over {T}}\hbox{$^{\lambda\mu}_{\sigma}\nabla_{\mu}\xi^{\sigma}$} 
=&(5.14)\cr
&=p^{a}_{A}e^{\lambda}_{a}\hbox{{\it \$}}_\Xi v^{A} + 
A^{\cdot\sigma\lambda}_{\rho}{\bar e}^{p}_{\sigma}
\hbox{{\it \$}}_\Xi e^{\rho}_{p} +
A^{\rho\lambda\sigma}\hbox{{\it \$}}_{\Xi} g_{\sigma\rho} + c.c. -
\lagAst\xi^{\lambda}\cr
W(L_{D},\Xi) &= 
{\buildrel D \over {W}}\hbox{$_{\sigma}\xi^{\sigma}$} 
+ {\buildrel D \over {W}}\hbox{$^{\mu}_{\sigma}\nabla_{\mu}\xi^{\sigma}$} 
=&(5.15)\cr
&=-H^{a}_{\mu}\hbox{{\it \$}}_\Xi e^{\mu}_{a} - 
e_{A}(L_{D})
\hbox{{\it \$}}_\Xi v^{A} + c.c. \cr
}$$
where
$$
\leqalignno{
A^{\rho\sigma\mu}&=p^{a}_{A}e^{\mu}_{a}\lambda^{A}_{Bij}
v^{B}{\bar e}^{i\rho}_{\ \cdot}{\bar e}^{j\sigma}_{\ \cdot}
&(5.16)\cr
H^{a}_{\mu}&=p^{a}_{\mu} + p^{a}_{A}\Omega^{A}_{\mu}
- \nabla_{\rho}A^{\cdot\sigma\rho}_{\mu}{\bar e}^{a}_{\sigma} - 
2 \nabla_{\sigma}A^{\sigma(\rho\nu)}{\bar e}^{a}_{\nu}g_{\mu\rho} + c.c. 
&(5.17)\cr
e_{A}(L_{D})&=p_{A} - e^{\mu}_{a}\nabla_{\mu}p^{a}_{A}
&(5.18)\cr
{\buildrel D \over {W}}\hbox{$_{\sigma}$}&=-e_{A}(L_{D})\Omega^{A}_{\sigma} + c.c. 
&(5.19)\cr
{\buildrel D \over {W}}\hbox{$^{\mu}_{\sigma}$}&= p^{a}_{\sigma}e^{\mu}_{a} + 
p^{a}_{A}e^{\mu}_{a}\Omega^{A}_{\sigma} - \nabla_{\rho}
(- A^{\mu\rho\lambda} + A^{\lambda\mu\rho} + A^{\rho\lambda\mu})
g_{\sigma\lambda} + c.c. 
&(5.20)\cr
{\buildrel D \over {T}}\hbox{$^{\lambda}_{\sigma}$}&=
p^{a}_{A} e^{\lambda}_{a}\Omega^{A}_{\sigma} + c.c.  - 
\lagAst\delta^{\lambda}_{\sigma}
&(5.21)\cr
{\buildrel D \over {T}}\hbox{$^{\lambda\mu}_{\sigma}$}&=
(A^{\mu\lambda\rho} - A^{\rho\mu\lambda} - A^{\lambda\rho\mu})
g_{\rho\sigma} + c.c. 
&(5.22)\cr
}$$

\noindent We remark once again that the contribution
to the vertical current vanishes identically.
The covariant conditions are therefore the following:

$$
\cases{
\nabla_{\lambda}{\buildrel D \over {T}}\hbox{$^{\lambda}_{\rho}$} + 
{1 \over 2}\,{\buildrel D \over {T}}\hbox{$^{\nu\mu}_{\sigma}$}
R^{\sigma}_{\rho\nu\mu} = {\buildrel D \over {W}}\hbox{$_{\rho}$} \cr
{\buildrel D \over {T}}\hbox{$^{\mu}_{\rho}$} + 
\nabla_{\lambda}{\buildrel D \over {T}}\hbox{$^{\lambda\mu}_{\rho}$}=
{\buildrel D \over {W}}\hbox{$^{\mu}_{\rho}$} \cr
{\buildrel D \over {T}}\hbox{$^{(\mu\nu)}_{\rho}$} = 0 \cr
}\nfor{5.23}$$
We can now define the {\sl total current} $E^{\lambda}(L,\Xi)$
and the {\sl total work} $W(L,\Xi)$ by setting:
$$
E^{\lambda}(L,\Xi)=E^{\lambda}(L_{H},\Xi) + E^{\lambda}(L_{D},\Xi)
\nfor{5.24}$$
$$
W(L,\Xi)=W(L_{H},\Xi) + W(L_{D},\Xi)
\nfor{5.25}$$
Once again we have:
$$
\nabla_{\lambda}E^{\lambda}(L,\Xi)=W(L,\Xi)
\nfor{5.26}$$
where:
$$\leqalignno{
E^{\lambda}(L,\Xi) &=
T^{\lambda}_{\sigma}\xi^{\sigma} 
+ T^{\lambda\mu}_{\sigma}\nabla_{\mu}\xi^{\sigma}
+ T^{\lambda\mu\nu}_{\sigma}\nabla_{\mu\nu}\xi^{\sigma}=&(5.27)\cr
&=({\buildrel H \over {T}}\hbox{$^{\lambda}_{\sigma}$} +
{\buildrel D \over {T}}\hbox{$^{\lambda}_{\sigma})\xi^{\sigma}$} 
+ ({\buildrel H \over {T}}\hbox{$^{\lambda\mu}_{\sigma}$} +
{\buildrel D \over {T}}\hbox{$^{\lambda\mu}_{\sigma})\nabla_{\mu}\xi^{\sigma}$} 
+ {\buildrel H \over {T}}
\hbox{$^{\lambda\mu\nu}_{\sigma}\nabla_{\mu\nu}\xi^{\sigma}$}\cr}$$
%\nfor{5.27}$$
$$
W(L,\Xi)=W_{\sigma}\xi^{\sigma} + W_{\sigma}^{\mu}\nabla_{\mu}\xi^{\sigma}=
{\buildrel H \over {W}}\hbox{$_{\sigma}\xi^{\sigma}$} + 
({\buildrel H \over {W}}\hbox{$^{\mu}_{\sigma}$} +
{\buildrel D \over {W}}\hbox{$^{\mu}_{\sigma})\nabla_{\mu}\xi^{\sigma}$}
\nfor{5.28}$$
and the total covariant conditions are the following:
$$
\cases{
\nabla_{\lambda}T^{\lambda}_{\rho} + 
{1 \over 2}\,T^{\nu\mu}_{\sigma}
R^{\sigma}_{\rho\nu\mu} +
{1 \over 3}\,\nabla_{\mu} R^{\sigma}_{\rho\lambda\nu} 
T^{\lambda\nu\mu}_{\sigma} = W_{\rho} \cr
T^{\mu}_{\rho} + 
\nabla_{\lambda}T^{\lambda\mu}_{\rho} + T^{\lambda\nu\mu}_{\sigma}
R^{\sigma}_{\rho\lambda\nu} -
{2 \over 3}\,
T^{\lambda\sigma\nu}_{\rho}
R^{\mu}_{\sigma\lambda\nu} = W^{\mu}_{\rho} \cr
T^{(\mu\nu)}_{\rho} +
\nabla_{\lambda}T^{\lambda\mu\nu}_{\rho} = 0 \cr
T^{(\lambda\mu\nu)}_{\rho} = 0 \cr}
\nfor{5.29}$$
The following weak conservation law is then valid on--shell
(i.e., along solutions of field equations) 
$$
\nabla_{\lambda}E^{\lambda}(L,\Xi)=d_{\lambda}E^{\lambda}(L,\Xi)=0
\nfor{5.30}$$
Hence $E(L,\Xi)=E^{\lambda}(L,\Xi)ds_{\lambda}$
is a conserved current on--shell, i.e.:
$$
E(L,\Xi, \rho) = (j^3\rho)^{\ast}E(L,\Xi)
\nfor{5.31}$$
is a closed form if and only if $\rho$ is a critical section.
Suitably manipulating on $(5.29)$ we get the so-called
{\sl generalised Bianchi identities}:
$$
W_{\sigma} - \nabla_{\mu} W_{\sigma}^{\mu} = 0 
\nfor{5.32}$$
Following the standard procedure of {\bf [14]} and {\bf [21]} define now a
$(m-2)$--form, called a {\sl superpotential}, by:
$$
U(L,\Xi) = {1 \over 2}\, \Big[\Big(T^{[\lambda\mu]}_{\rho} - {2 \over 3}\,
\nabla_{\sigma} T^{[\lambda\mu]\sigma}_{\rho}\Big)\xi^{\rho} + 
\Big({4 \over 3}\,T^{[\lambda\mu]\nu}_{\rho}\Big)\nabla_{\nu}\xi^{\rho}\Big]
\, ds_{\lambda\mu}
\nfor{5.33}$$
and a $(m-1)$--form ${\tilde E}(L,\Xi)$, called {\sl the reduced energy-density},
by:
$$
{\tilde E}(L,\Xi) = W_{\sigma}^{\mu}\xi^{\sigma} ds_{\mu}
\nfor{5.34}$$
For the energy-density flow the following representation is thence true
$$
E(L,\Xi) = {\tilde E}(L,\Xi) + \hbox{Div}\, U(L,\Xi)
\nfor{5.35}$$
where Div denotes the formal divergence, defined
for any global section $\rho$ and any $p$--form
$\omega$ over $J^{2}B$ by:
$$
(j^{3}\rho)^{\ast}(\hbox{Div}\,\omega) = 
d \Big[ (j^{2}\rho)^{\ast}\omega \Big] 
\nfor{5.36}$$
We finally consider the global forms $U(L,\Xi, \rho)=(j^2\rho)^{\ast}U(L,\Xi)$
and  ${\tilde E}(L,\Xi, \rho)=(j^3\rho)^{\ast}{\tilde E}(L,\Xi)$
obtained by pull back along any section $\rho$; the second one vanishes on shell
(i.e. if $\rho$ is critical).
Therefore, the energy-density flow $E(L,\Xi)$ is an exact form
along critical sections.

We remark that these results for conserved currents are completely general and they hold
actually for every covariant Lagrangian.
If we turn back to our choice $(3.4)$ and $(3.5)$ of the total Lagrangian we get the
following explicit expression for the superpotential:
$$
\eqalign{
U(L,\Xi):&=U(L_{H},\Xi)+U(L_{D},\Xi)\cr
U(L_{H},\Xi)&=  {1\over 2}{1\over
2\kappa}\Big[\na^\mu\xi^\nu-\na^\nu\xi^\mu\Big]\sqrt{g}\>ds_{\mu\nu}\cr U(L_{D},\Xi)&=  {i\over
8}\bar v\Big[(\ga^{[\mu}\ga^{\nu]}\ga^{\rho} +2g^{\rho[\mu}\ga^{\nu]})\xi_\rho\Big]v\sqrt{g}
\>ds_{\mu\nu}\cr
}\leqno{(5.37)}$$
where we have set $\ga^{\mu}=e^\mu_a\ga^{a}$.

Since $W(L,\Xi)$ vanishes on shell, pulling back $U(L,\Xi)$ on a solution and integrating
it on the border of a spatial domain $D$, one gets a conserved quantity for each
1-parameter family of symmetries:
$$
Q_\Xi:=\int_{\de D} (j^1\rho)^\ast U(L,\Xi)
\leqno{(5.38)}$$
\section{VI Conclusion and perspectives}

The new formalism we have developed for spinor theories is interesting for
some further reasons besides those we already mentioned in the Introduction.

First of all we see once again that the requirement of geometric coherence
actually allows us to select the theory we are looking for within the much larger
set of possible theories.
In fact here, as well as in our previous work concerning Bosonic matter \nP2,
we have not been concerned with the existence of solutions, but we
have just cared that all concepts entering the theory could be well defined by a global and
geometric point of view. The surprise arises from the fact that even under
these general and formal requests we can manage to build {\it physically admissible} theories and
above all to set aside lots of other theories.

The second reason is related to researches about a unifying paradigm for General
Relativity and Quantum Mechanics. Here it is important to fix what
spinor matter is in General Relativity, since the new paradigm must reproduce
this formalism in the classical limit (= not Quantum) and spinors of Quantum Mechanics
in the flat limit (= on Minkowski space).
However since spinor theories on curved spaces are qualitatively much more
complicated, we can perhaps expect that the first request is stronger than
the second one.

We finally want to stress which are the differences between the classical spinor
theories on curved spaces (see, e.g., \nOla) and our approach.
First of all, our approach is naturally formulated in the framework of
variational calculus on fibred bundles.
Second, our formalism allows a deep analogy among spinor theories,
General Relativity as formulated here, so called {\it natural theories} and gauge
theories as formulated in \nP2\ and \nPartI. In each case we start by choosing
a principal bundle (called {\sl structure bundle}); the configuration bundle
is then a gauge-natural bundle associated to some principal prolongation of
it and the Lagrangian
is required to be covariant with respect to any principal automorphism of the 
structure bundle represented on the configuration bundle.
In the case of General Relativity this result is obtained in a way that
we believe is important to stress.
We have both enlarged the symmetry group from Diff($M$) to $\Aut{\Sigma}$
and the number of dynamical fields from $g_{\mu\nu}$ to $e_a^\mu$.
The covariance request on the Lagrangian lets $e_a^\mu$ appear when
we are coupling with spinor matter but allows just $g_{\mu\nu}$
when coupling with Bosonic matter.
Further investigations about the unified formulation of
the above theories should therefore follow 
and we hope to address this problem in the near future.

\section{VII Acknowledgements}

We are deeply grateful to A. Borowiec for useful discussions about the
foundations of the theory.
This work is sponsored by G.N.F.M.--C.N.R. and by M.U.R.S.T. (Nat. Proj.
``Met. Geom. e Probab. in Fisica Matematica'').
\section{Appendix}

The following Lie derivatives are easily calculated:

$$
\leqalignno{
\hbox{{\it \$}}_\Xi e^{\mu}_{a}&=
e^{\mu}_{b}(\xi_{V})^{b\cdot}_{\, a} - e^{\nu}_{a}\nabla_{\nu}\xi^{\mu}
&(A.1) \cr
\hbox{{\it \$}}_\Xi v^{A} &= \Omega^{A}_{a}{\bar e}^{a}_{\mu}\xi^{\mu} - 
\lambda^{A}_{Bij} v^{B} (\xi_{V})^{ij}
&(A.2a) \cr
(\xi_{V})^{ij}&:= \xi^{ij} - \Gamma^{ij}_{\,\,\mu}\xi^{\mu}
&(A.2b) \cr
\hbox{{\it \$}}_\Xi \Omega^{A}_{a} &= \xi^{\mu}\nabla_{\mu}\Omega^{A}_{a} +
(\xi_{V})^{b\cdot}_{\, a}\Omega^{A}_{b} -
\lambda^{A}_{Bij}\Omega^{B}_{a}(\xi_{V})^{ij}
&(A.3)\cr
\hbox{{\it \$}}_\Xi \Gamma^{ij}_{\,\,\mu} &= R^{ij}_{\,\,\rho\mu}\xi^{\rho} +
\nabla_{\mu}(\xi_{V})^{ij}
&(A.4)\cr
\hbox{{\it \$}}_\Xi \Gamma^{\rho}_{\sigma\mu} &= 
\hbox{{\it \$}}_\xi \Gamma^{\rho}_{\sigma\mu} = 
- R^{\rho}_{\,(\sigma\mu)\nu}\xi^{\nu} +
\nabla_{(\sigma}\nabla_{\mu)}\xi^{\rho}
&(A.5)\cr
\hbox{{\it \$}}_\Xi g_{\mu\nu} &= \hbox{{\it \$}}_\xi g_{\mu\nu} = 
\nabla_{\mu}\xi^{\sigma} g_{\sigma\nu} + \nabla_{\nu}\xi^{\sigma} g_{\sigma\mu}
&(A.6)\cr
}$$

Moreover it is not difficult to prove the following identities:
$$
\leqalignno{
\hbox{{\it \$}}_\Xi \Omega^{A}_{a} &= \Omega^{A}_{b}{\bar e}^{b}_{\mu} 
\hbox{{\it \$}}_\Xi e^{\mu}_{a} + 
e^{\mu}_{a}(d_{\mu}\hbox{{\it \$}}_\Xi v^{A} + \lambda^{A}_{Bij}
\Gamma^{ij}_{\,\,\mu}\hbox{{\it \$}}_\Xi v^{B} + 
\lambda^{A}_{Bij} v^{B} \hbox{{\it \$}}_\Xi \Gamma^{ij}_{\,\,\mu})
&(A.7)\cr
\hbox{{\it \$}}_\Xi \Gamma^{ij}_{\,\,\mu} &= {\bar e}^{i}_{\rho}
(\hbox{{\it \$}}_\xi \Gamma^{\rho}_{\sigma\mu} +
{\bar e}^{k}_{\sigma}\nabla_{\mu}\hbox{{\it \$}}_\Xi e^{\rho}_{k})
e^{\sigma}_{p}\eta^{pj}
&(A.8)\cr
}$$

We finally observe that:
$$
\leqalignno{
\nabla_{\mu}e^{\nu}_{a}&:= d_{\mu}e^{\nu}_{a}
+ e^{\sigma}_{a}\Gamma^{\nu}_{\sigma\mu}
- \Gamma^{b}_{\,a\mu}e^{\nu}_{b} \equiv 0
&(A.9) \cr
}$$

\section{References}

\noindent\nP2\ \titP2

\noindent\nHae\ \titHae

\noindent\nBorHi\ \titBorHi

\noindent\nMil\ \titMil

\noindent\nLawMi\ \titLawMi

\noindent\nGP\ \titGP

\noindent\nPenone\ \titPenone

\noindent\nOla\ \titOla

\noindent\nGe\ \titGe

\noindent\nBW\ \titBW

\noindent\nPentwo\ \titPentwo

\noindent\nHE\ \titHE

\noindent\nF3G\ \titF3G

\noindent\nKos\ \titKos

\noindent\nRoot\ \titRoot

\noindent\nPartI\ \titPartI

\noindent\nCamI\ \titCamI

\noindent\nCamII\ \titCamII

\noindent\nStress\ \titStress

\noindent\nKol\ \titKol

\noindent\nGeo\ \titGeo

\noindent\nGau\ \titGau

\noindent\nNatI\ \titNatI

\noindent\nNatII\ \titNatII

\noindent\nRob\ \titRob

\end